\documentclass[12pt,here]{article}

\usepackage{mathtext}
\usepackage{amssymb,amsmath}
\usepackage{epsfig}
\usepackage{fullpage}
\usepackage{cite}
\usepackage{delarray} 
\usepackage{lscape}

\newcommand{\kpm}   {\mbox{$K^+ \! \to \! \pi^+ \mu^+ e^-$}}
\newcommand{\kk}    {\mbox{$K^0 \! \rightleftarrows \! \bar{K}^0$}}
\newcommand{\kpme}  {\mbox{$K^+ \! \to \! \pi^+\mu^- e^+$}}
\newcommand{\kem}   {\mbox{$K^0_L \to e^- \mu^+$}}
\newcommand{\klem}  {\mbox{$K^0_L \to e^\mp \mu^\pm$}}
\newcommand{\klpem} {\mbox{$K^0_L \to\pi^0 e^\mp \mu^\pm$}}
\newcommand{\kmn}   {\mbox{$K^+ \to \mu^+ \nu_\mu$}}

\newcommand{\kXk}   {\mbox{$K^0\rightleftarrows X^0\rightleftarrows\bar{K}^0$}}
\newcommand{\kpmm}  {\mbox{$K^+ \to \pi^-\mu^+\mu^+$}}
\newcommand{\piHwK}  {\mbox{$\langle\pi|H_w|K\rangle$}}
\newcommand{\HwK}    {\mbox{$\langle 0|H_w|K^0_L\rangle$}}

\def\BR{\mathrm{BR}}

\begin{document}

\newlength{\vlength}
\vlength=\baselineskip
\def\tov{%
    \kern0.5em
    \hbox{%
        \vtop to 0pt{%
            \vss
            \hbox{\vrule height0.8\vlength depth-0.17\vlength}
        }
    }
    \kern-0.70em
    \to}
\def\tovv{%
    \kern0.5em
    \hbox{%
        \vtop to 0pt{%
            \vss
            \hbox{\vrule height1.8\vlength depth-0.17\vlength}
        }
    }
    \kern-0.66em
    \to}
\def\tovvv{%
    \kern0.5em
    \hbox{%
        \vtop to 0pt{%
            \vss
            \hbox{\vrule height2.8\vlength depth-0.17\vlength}
        }
    }
    \kern-0.66em
    \to}

\author
{L.G.~Landsberg \\
\small {Institute for High Energy Physics, Protvino, Russia}}

\title{
\vskip -1.5cm
\rightline{\small{IHEP Preprint 2004-33}}
\vskip 1.5cm
Is it still worth searching for lepton flavor violation in rare kaon decays?
\thanks{An extended version of the talk given at the Chicago Flavor Seminar, February 27, 2004.}
}

\date{}

\maketitle

\begin{abstract}

Prospective searches for lepton flavor violation (LFV) in rare kaon
decays at the existing and future intermediate-energy accelerators 
are considered. The proposed studies are complementary to LFV searches 
in muon-decay experiments and offer a unique opportunity to probe 
models with approximately conserved fermion-generation quantum number 
with sensitivity superior to that in other processes. Consequently, 
new searches for LFV in kaon decays are an important and independent 
part of the general program of searches for lepton flavor violation 
in the final states with charged leptons.

\end{abstract}

\section{Fundamental fermion generations of the Standard Model, lepton
flavors, and neutrino oscillations}

The Standard Model (SM) provides a good description of physics
phenomena in the range of masses up to several hundred GeV. The SM 
includes three generations of fundamental particles -- quarks and leptons:

\begin{eqnarray}
u, d, e, \nu_e & - & \mbox{first generation}, \nonumber \\
c, s, \mu, \nu_{\mu} & - & \mbox{second generation}, \nonumber \\
t, b, \tau, \nu_{\tau} & - & \mbox{third generation}.
\end{eqnarray}

Strong interactions between quarks (characterized by special quantum
numbers: flavors and colors) are realized by the exchange of eight types 
of massless colored vector gluons. These interactions are described 
within the framework of the modern theory of strong processes --
Quantum ChromoDynamics (QCD). QCD formalism allows to perform reliable 
perturbative calculations at the leading or next-to-leading logarithmic 
order at sufficiently short distances (below $1-2\,\mbox{GeV}^{-1}$).
The fact that quark and gluons carry color quantum number makes them 
unobservable as free particles (the concept of confinement). Strong 
interactions conserve quark flavor; consequently quarks cannot change 
flavor in strong processes and can only be rearranged into various combinations, 
produced in pairs ($q\bar{q}$), or undergo annihilation of such pairs.

Electroweak interactions involving quarks and leptons are carried by 
the massive intermediate vector bosons $W^{\pm}$, $Z$ and the massless 
photon $\gamma$. Weak interactions of charged currents carried by the 
$W^{\pm}$-bosons change the quark flavors. (A scalar Higgs boson $H^0$ 
predicted in the SM is also expected to play an important role in weak 
interactions; however its very existence is yet to be established experimentally.)

The Lagrangian of electroweak interaction is based on the broken 
$SU(2)_L \times U(1)_Y$ group, which includes left-handed quarks and 
leptons that form weak-isospin doublets:
\begin{eqnarray}
\left( \begin{array}{c}
\nu_e \\ e^- \end{array} \right)_L, & & 
\left( \begin{array}{c}
\nu_{\mu} \\ \mu^- \end{array} \right)_L. \quad
\left( \begin{array}{c}
\nu_{\tau} \\ \tau^- \end{array} \right)_L, 
\end{eqnarray}
\begin{eqnarray}
\left( \begin{array}{c}
u \\ d^{'} \end{array} \right)_L, & &
\left( \begin{array}{c}
c \\ s^{'} \end{array} \right)_L, \qquad
\left( \begin{array}{c}
t \\ b^{'} \end{array} \right)_L, 
\end{eqnarray}
as well as right-handed quarks and leptons $q_R,l_R$, that form weak-isospin 
singlets. Here d$^{'}$, s$^{'}$, and b$^{'}$ are mixed quark-states, with 
mixing described by the CKM-matrix. 

The left and right-handed fermions can be represented as 
\begin{equation}
\psi_L = \frac{1}{2}(1-\gamma_5)\psi, \quad \quad 
\psi_R = \frac{1}{2}(1+\gamma_5)\psi.
\end{equation}

Leptons of the fundamental fermion generations are characterized by 
the lepton flavors ($L_e$ for the first generation, $L_\mu$ for the 
second one, and $L_\tau$ for the third  family). In the past it was 
assumed that all these lepton flavors are conserved and that all the 
neutrinos are massless particles. We also will use $L=L_e+L_{\mu}+L_\tau$ -- total 
lepton number. 

However, lepton flavor conservation is not necessary forced by a global 
symmetry of a $U(1)$-type, as is the case, e.g., for the conservation 
of electric charge. Therefore, searches for lepton flavor violation (LFV) constitute an important direction in experimental particle physics and, as such, have been carried out for the past half-a-century. Despite of numerous experiments, no LFV-processes involving charged leptons have been observed as of yet, while the sensitivity of these searches has been steadily increasing by about two orders of magnitude every decade (see Fig.~1 [1]). Current upper limits on branching fractions of various LFV-decays with charged leptons are 
presented in Table 1 [2--12].

It was a brilliant idea by Bruno Pontecorvo [13] of the possibility of 
neutrino oscillations if lepton flavors are not conserved, which, after 
four decades of experimental and theoretical efforts, has led to a great 
discovery. Neutrino oscillations have been observed in experiments with 
atmospheric and solar neutrinos [14,15], reactor neutrinos [16], and possibly
even accelerator neutrinos [17] (see also recent reviews [18]).

It is worth emphasizing that no radical changes to the SM are required to 
explain the neutrino oscillations. Apparently it is sufficient to 
make only a minor change in the SM Lagrangian by including a neutrino 
mass-term that is non-diagonal in flavor neutrino fields. 
In addition, a neutrino mixing matrix must be introduced; 
this PMNS-matrix (Pontecorvo-Maki-Nakagava-Sakata [13,19]) is somewhat similar to the 
CKM-matrix that describes quark mixing. However, while the CKM-matrix is 
responsible for rich physics in the quark sector of the SM, the PMNS-matrix may result 
only in quite limited phenomenology.  

\begin{figure}
\centering \includegraphics[height=10cm,width=10cm]{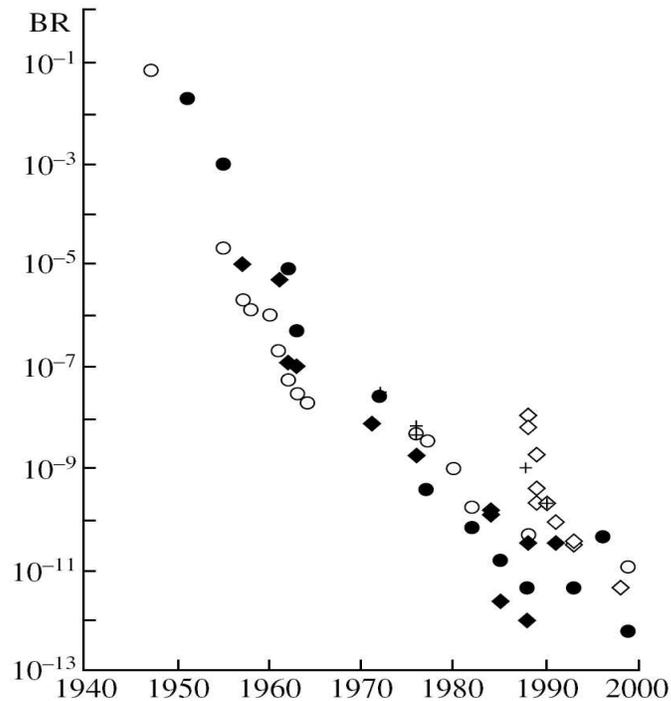}
\caption{Historical progress of LFV-searches for various processes
with muons and kaons [1]. Notations: $\circ$ - $\mu \to e \gamma$, 
$\blacklozenge$ - $\mu \to 3e$, $\bullet$ - $\mu^- + A \to e^- + A$ 
-- muon processes; $\lozenge$ - $K^0_L \to e \bar{\mu}$, $+$ - 
$K^+ \to \pi^+ e \bar{\mu}$ -- kaon processes. As seen from the plot, 
the sensitivity of LFV searches has been increasing on average by two orders of 
magnitude per decade.
}
\end{figure}
\begin{figure}
\centering \includegraphics[height=6cm,width=12cm]{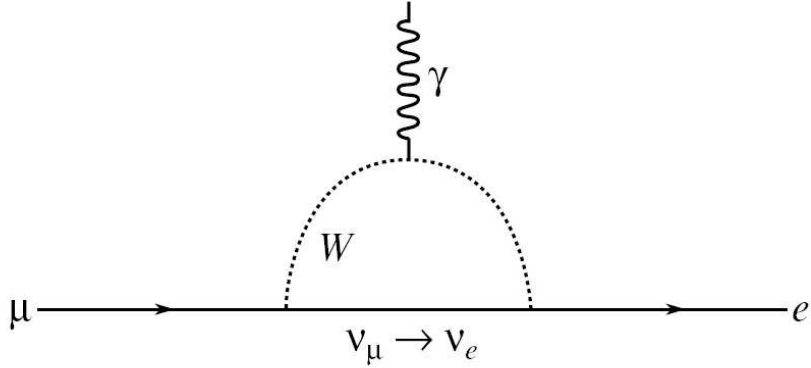}
\caption{The diagram of the $\mu \to e \gamma$ decay via neutrino mixing.}
\end{figure}

To illustrate this limitation, let us consider other LFV-processes with charged 
leptons, e.g., a LFV decay $\mu\to e + \gamma$. As it might appear from examining the Feynman diagram in Fig.~2, this decay could be explained by the neutrino-mixing mechanism. However, the fact that neutrino masses are so small suppresses the branching fraction of this decay way below our ability to detect it experimentally (see [20]): 

\begin{equation}
\label{math/5}
\BR(\mu\to e\gamma) = \frac{\Gamma(\mu\to e\gamma)}{\Gamma(\mu\to e\bar{\nu}_e\nu_\mu)} \approx 
\frac{3}{32}\cdot
\frac{\alpha}{\pi}\left|\sum_iU^*_{\mu i}U_{ei}\left(\frac{m_{\nu i}}{M_W}\right)^2\right|^2\lesssim 10^{-48}.
\end{equation}
Here $U$ is the PMNS-matrix, $\nu_i$ are neutrino mass and lepton flavor (LF) eigenstates ($m_{\nu_i}<1$ eV), and $\nu_\mu = \sum_i U_{\mu i} \nu_i,$ $\nu_e = \sum_i U_{ei} \nu_i$ are the states produced in association with muons and electrons; they are neither mass nor LF eigenstates.

The above example illustrates that it is possible that neutrino oscillations 
are the only processes with observable LFV. If neutrinos are Majorana particles 
the other observable new effect can be a neutrinoless double-$\beta$-decay
$Z\to (Z+2)+2e^-$ with electron lepton number violation, $|\Delta L_e| = 2$.

Nevertheless, we hope for a more interesting possibility: a discovery of
new physics phenomena beyond the SM in LFV decays ($\mu\to e\gamma$, 
$\mu\to 3e$, $K^0_L\to e\bar{\mu}$, $K^+\to\pi^+\mu^+ e^-$, etc). These
searches have been a subject of growing interest from the particle physics 
community, especially after the observation of LFV in neutrino oscillations. 
The latter have established that lepton numbers $L_e$, $L_\mu$, and  
$L_\tau$ are not conserved separately. There are many theories beyond the SM 
that predict such a violation in the processes with charged leptons, which 
are forbidden in the SM. As a rule (see below), many of these new LFV mechanisms are sensitive to very high energy-scales that cannot be probed directly even with the new generation of supercolliders. Thus, searches for rare LFV decays may offer a unique window to this type of new physics.

As there are different types of LFV-processes, they may be sensitive to
different mechanisms of lepton flavor violation and thus yield complementary 
information about its origin:

\begin{itemize}
\item[a)] Pure leptonic LFV-processes ($\mu\to e\gamma$,
 $\mu\to 3e$, $\tau\to\mu\gamma$, 
$\tau\to e\gamma$, $\tau\to 3l$, etc).
\item[b)] Quark-lepton LFV processes of the $d\to d\mu\bar{e}$ type
 (neutrinoless conversion $\mu^- + (A,Z) \to e^- + (A,Z)$).
\item[c)] Quark-lepton LFV processes of 
 $s\to d\mu\bar{e}$ type (kaon LFV decays $K^0_L\to \mu\bar{e}$,
 $K\to\pi\mu \bar{e}$, etc). 
\item[d)] Lepton-number-violating decays -- $L$-nonconservation (neutrinoless 2$\beta$-decays, 
$K^+ \to \pi^- l^+ l^+$, etc).
\end{itemize} 

In principle all these four types of processes are important for future
searches for lepton flavor violation. However, to date only the first 
two types of processes (and neutrinoless double-$\beta$-decay) have been 
subject of the renewed theoretical and experimental interest. Several
ambitious projects have been proposed to search for the $\mu\to e + \gamma$ decay up to the branching fraction of $\sim 10^{-14}$ [21], which may be 
further lowered by one or two orders of magnitude at future neutrino 
factories (similar to the case of the $\mu\to 3e$ searches). Using the 
idea of Ref. [22] and the new technique of superconductive traps, the MECO 
experiment has proposed to increase the sensitivity  of $F = \frac{\mu^- 
+ (A,Z)]\to e^- + (A,Z)}{(\mu^- + A,Z\to \mbox{capture})}$ to the level of 
$10^{-17}$ [23], which may be further improved to $10^{-19}$ by the 
J-PARC project [24]. 

On the other hand, there are no specific new proposals and quite limited 
discussion on further searches for LFV kaon decays. Nevertheless, as we 
show below, these processes may possess certain unique properties and could 
provide complementary information about lepton flavor violation compared 
to the better-explored processes a) and b).

\small
\begin{center}
\begin{table}[t]
\caption{Upper limits for BR of the LFV processes [2-12]}
\begin{center}
\footnotesize
\begin{tabular}{|l|c|c|} \hline
      &  &  \\ 
 & 90\% C.L. upper limit on BR & Future proposals  \\ 
        &  &  \\ \hline
$\klem$ & $4.7\times 10^{-12}$ (BNL E871) [2] &  \\ \hline
$\klpem$ & $3.3\times 10^{-10}$ (KTeV) [3] &  \\ \hline
$K^0_L\to e^{\pm}e^{\pm}\mu^\mp\mu^\mp$ & $4.12\times 10^{-11}$ (KTeV) [3,4]  &   \\ \hline
$\kpm$ & $1.2\times 10^{-11}$ (BNL 865) [5,10] &  \\ \cline{1-2}
$\kpme$ & $5.2\times 10^{-10}$ (BNL 865) [6] & CKM  \\ \cline{1-2}
$K^+\to\pi^-e^+e^+$ & $6.4\times 10^{-10}$ (BNL 865) [6] & $\sim 10^{-12}$  \\ \cline{1-2}
$K^+\to\pi^-e^+\mu^+$ & $5.0\times 10^{-10}$ (BNL 865) [6] &  \\ \cline{1-2}
$K^+\to\pi^-\mu^+\mu^+$ & $3.0\times 10^{-9}$ (BNL 865) [6] &   \\ \cline{1-2}
$\Xi^-\to p\mu^-\mu^-$  & $<4.0\times 10^{-8}$ [11] & \\ \hline
$\mu\to e\gamma$ & $1.2\times 10^{-11}$ (MEGA) [7] & $10^{-14}$ (PSI)
[21]; \hspace{10mm} neutrino factories
 \\ \cline{1-2} 
$\mu\to 3e$ & $1.0\times 10^{-12}$ (SINDRUM) [8] & \hspace{30mm} $\sim 10^{-15}$ \\ \hline
$\mu^- + Ti \to e^- + Ti$ & $4.3\times 10^{-12}$ (SINDRUM II) [9] &
$10^{-17}$ (MECO) [23]; $10^{-19}$ J-PARC [24] \\ \hline
$\tau\to e\gamma,\mu\gamma,3l$ & $\lesssim  10^{-6} - 10^{-7}$ [12] &  \\ \hline
\end{tabular}
\end{center}
\end{table}
\end{center}
\normalsize

\section{Phenomenology of kaon LFV-decays}

Let us consider kaon decays with lepton flavor violation:

\begin{equation}
\label{math/6}
K^0_L \to e^-\mu^+
\end{equation}
and
\begin{equation}
\label{math/7}
K\to\pi e^-\mu^+.
\end{equation} 
Here we present a brief phenomenological description of these processes
[25,26] (see also Refs. [27,28]). Only axial and pseudoscalar hadron currents can contribute to the \HwK\ amplitude of the decay (6), since the $K$-meson is a pseudoscalar particle. On the other hand, scalar, vector, and tensor hadron currents can contribute to the \piHwK\ amplitude of decay (7), since the $\pi$ meson is also a pseudoscalar. The matrix element for the decay \kem\ is:

\begin{equation}
\label{math/8}
M  =  \frac{G_F}{\sqrt{2}}\left[J_A^\lambda(f'_A\bar{u}_e\gamma_\lambda v_\mu + f_A\bar{u}_e\gamma_5
\gamma_\lambda v_\mu)  +  J_P(f'_P\bar{u}_ev_\mu + f_P\bar{u}_e\gamma_5v_\mu)\right],
\end{equation}
with the axial and pseudoscalar hadron currents given by:

\begin{eqnarray}
\label{math/9}
J^\lambda_A & = & P^\lambda m_Ka_A\varphi_K = P^\lambda m_Ka_A\frac{1}{\sqrt{2m_K}}~\mbox{~and} \nonumber \\
J_P & = & m^2_Ka_P\varphi_K = m_K^2 a_P \frac{1}{\sqrt{2m_K}},
\end{eqnarray}
respectively. Here $\varphi_K$ is the pseudoscalar kaon wave function with the normalization
 $\frac{1}{\sqrt{2m_K}}$,
$P^\lambda$ is the four-momentum of the kaon ($P^\lambda = P_\mu^\lambda +
P^\lambda_e$), and $a_P$, $a_A$ are dimensionless parameters determined by the means of current algebra from the comparison with the $K_{\mu_2}$ decay with the kaon decay constant $f_K = 159.8\pm 1.5$ MeV: $a_A = \frac{\sqrt{2}f_K}{m_K} = 0.46$ and $a_P = a_A \frac{m_K}{m_s + m_d} = 2.1$, where $m_s \approx 100$ MeV and $m_d\approx 7.5$ MeV are the strange and down-quark ``current'' masses [25]. The terms $\frac{G}{\sqrt{2}}f_A$, $\frac{G}{\sqrt{2}}f'_A$,
 $\frac{G}{\sqrt{2}}f_P$, and $\frac{G}{\sqrt{2}}f'_P$ describe general lepton flavor violating interactions of the kaon and non-SM lepton currents. 
Since the LFV interactions are already accounted for in Eq. (8), we do not include the Cabibbo angle $\sin\vartheta_c$ in the expression for the matrix element $M$.

Using Dirac equation for lepton spinors $(\hat{P}_\mu + m_\mu)v_\mu
 = (P_\mu^\lambda\gamma_\lambda + m_\mu)v_\mu = 0$ 
and $\bar{u}_e(\hat{P}_e - m_e) = 0$ (where $v_\mu$ and $u_e$ are spinors representing the antiparticle $\mu^+$ and the particle $e^-$) and the commutative properties of the Dirac $\gamma$-matrices, we obtain:

\begin{eqnarray}
\label{math/10}
M & =  & \frac{G_F}{\sqrt{2}}\left\{\frac{m_Ka_K}{\sqrt{2m_K}}\left[f'_A \bar{u}_e(\hat{P}_e + 
\hat{P}_\mu)v_\mu +
f_A \bar{u}_e\gamma_5(\hat{P}_\mu + \hat{P}_e)v_\mu\right] + \right. \nonumber \\
&  & \left.\frac{m_K^2a_P}{\sqrt{2m_K}}\left[f'_P (\bar{u}_ev_\mu) +
f_P (\bar{u}_e\gamma_5v_\mu)\right] \right\} = \nonumber \\
& & \frac{G_F}{\sqrt{2}}\left\{\frac{m_Ka_K}{\sqrt{2m_K}}\left[f'_A (m_e\bar{u}_ev_\mu - m_\mu\bar{u}_e
v_\mu) +
f_A (-m_e\bar{u}_e\gamma_5v_\mu - m_\mu\bar{u}_e\gamma_5v_\mu)\right] + \right. \nonumber \\
& & \left. \frac{m_K^2a_P}{\sqrt{2m_K}}\left[f'_P\bar{u}_ev_\mu) +
f_P\bar{u}_e\gamma_5v_\mu\right]\right\} 
 =  \frac{1}{\sqrt{2m_K}}\left[A\bar{u}_e\gamma_5v_\mu + B\bar{u}_ev_\mu\right],
\end{eqnarray}
with the dimensionless amplitudes $A$ and $B$ given by:

\begin{equation}
\label{math/11}
\left.\begin{array}{ccl}
A & = & \displaystyle\frac{\mathstrut G_F}{\sqrt{2}}m_K \left[-f_A (m_\mu + m_e)a_A + 
f_Pm_Ka_P\right] \approx \\
 & & \displaystyle\frac{\mathstrut G_F}{\sqrt{2}}m_Km_\mu a_A \left[-f_A+ f_P 
\displaystyle\frac{\mathstrut m_Ka_P}{m_\mu a_A}\right] \approx 2.00\times 10^{-7}[-f_A + 9.9f_P] \\ 
B & = &  \displaystyle\frac{\mathstrut G_F}{\sqrt{2}}m_Ka_A \left[
-f'_A (m_\mu - m_e)a_A + f'_Pm_K\frac{a_P}{a_A}\right] \approx \\  
 & & \displaystyle\frac{\mathstrut G_F}{\sqrt{2}}m_Km_\mu 
a_A \left[-f'_A+ f'_P
\displaystyle\frac{\mathstrut m_Ka_P}{m_\mu a_A}\right] \approx 2.00 \times 10^{-7}[-f'_A + 9.9f'_P] 
\end{array}\right\}.
\end{equation}
The square of the matrix element can be found using standard techniques 
(see, e.g., Ref. [29]) to be:

\begin{eqnarray}
\label{math/12}
|M|^2 & =  & 
 \frac{4}{2m_K}\left[|A|^2 + |B|^2\right](P_\mu P_e) +
 \frac{4}{2m_K}\left[|A|^2 - |B|^2\right](m_\mu m_e) \approx \nonumber \\
& &  \frac{4}{2m_K}\left[|A|^2 + |B|^2\right](P_\mu P_e)
\end{eqnarray}
(here we neglected the term proportional to $m_e$), which corresponds to the decay (6) width of:

\begin{eqnarray}
\label{math/13}
\Gamma(\kem) & = &  \frac{1}{(2\pi)^2}\int\frac{{\rm d}^3\vec{P}_\mu}{2E_\mu} \frac{{\rm d}^3\vec{P}_e}{2E_e}
  |M|^2\delta^{(4)}(P_K + P_\mu + P_e) = \nonumber \\
& & \frac{m_K}{8\pi}\left(1-\frac{m_\mu^2}{m^2_K}\right)^2[|A|^2 + |B|^2] \approx 18.1\,\mbox{MeV} \times [|A|^2 + |B|^2].
\end{eqnarray}

In models with $|f_A| = |f'_A|$ and $f_P = f'_P =0$ (axial-only interaction) we obtain from Eq. (11) and $a^2_A = 2f^2_K/m^2_K$ (with kaon decay constant $f_K = 159.8\pm 1.5$~MeV):

\begin{equation}
|A|^2+|B|^2 = 2G_F^2|f_A|^2f^2_K m_\mu^2,
\end{equation}
\begin{equation}
\label{math/15}
\BR(\kem)\cdot\tau (K^0_L)^{-1} =  \Gamma(\kem) = \frac{2G_F^2|f_A|^2m_K}{8\pi}f_K^2m_\mu^2\left(1-\frac
{m_\mu^2}{m_K^2}\right)^2
\end{equation}
and
\begin{equation}
\label{math/16}
\BR(\klem) \tau (K^0_L)^{-1}  =  \Gamma(\klem) = \frac{4G_F^2|f_A|^2m_K}{8\pi}f_K^2m_\mu^2\left(1-\frac
{m_\mu^2}{m_K^2}\right)^2.
\end{equation}
Let's compare this LFV kaon decay with the SM $\kmn$ decay (see, e.g.,  Ref. [29]):

\begin{equation}
\label{math/17}
\BR(\kmn)\tau (K^+)^{-1} = \Gamma(\kmn) = \frac{G_F^2\sin^2\vartheta_cm_K}{8\pi}f_K^2m_\mu^2\left(1-\frac
{m_\mu^2}{m_K^2}\right)^2.
\end{equation}
From Eqs. (16), (17), Table 1, and Ref. [12] we find, at the 90\% confidence level (C.L.):

\begin{equation}
\label{math/18}
B_1 = \left[\frac{\BR(\klem)}{\BR(\kmn)}\frac{\tau (K^+)}{\tau (K^0_L)}\right]  = 
\frac{4|f_A|^2(G_F/\sqrt{2})^2}{\sin^2\vartheta_c (G_F/\sqrt{2})^2} \leq 
1.75 \times 10^{-12}\; (90\%\,\mbox{C.L.}).
\end{equation}
Let's consider a dynamical model for the decay (6) with an $s$-channel exchange of a new boson, $X^0$, which couples to the $e^{\mp}\mu^{\pm}$ pairs (see Feynman diagram in Fig.~3). In this case the probability of the decay (6) is 
proportional to

\begin{equation}
\label{math/19}
|f_A|^2\frac{G_F^2}{2} = \left(\frac{h'h''}{M_X^2}\right)^2
\end{equation}
and, from Eqs. (\ref{math/18}) and (\ref{math/19}),

\begin{eqnarray}
\label{math/20}
B_1 & = & \frac{4\left(\frac{h'h''}{M_X^2}\right)^2}{\sin^2\vartheta_c\left(\frac{g^2}{8M^2_W}\right)^2} = 
\frac{4}{\sin^2\vartheta_c}\frac{\left(\frac{h'h''}{M_X^2}\right)^2}{\left(\frac{g^2}{8M_W^2}\right)^2} = 
4\left(\frac{M_W}{M_X}\right)^4\left[\left(\frac{h'h''}{g^2/8}\right)^2\frac{1}{\sin^2\vartheta_c}\right] =
\nonumber \\
& & \left[\frac{16}{(\sin\vartheta_c g^2)}\right]^2\left(\frac{h'}{h''}\right)^2\left[
\frac{h''}{M_X}M_W\right]^4 =
2.86\times 10^4\left(\frac{h'}{h''}\right)^2\left[\frac{h''}{M_X}M_W\right]^4.
\end{eqnarray}

The mass of the $X^0$ particle can be obtained from Eq. (\ref{math/20}):
\begin{equation}
M_X^4 = \frac{4M_W^4}{B_1}\left[\left(\frac{h'h''}{g^2/8}\right)^2\frac{1}{\sin^2\vartheta_c}\right].
\end{equation}
Assuming $\left(\frac{h'h''}{g^2/8}\right)^2 \approx 1$, we find

\begin{equation}
\label{math/22}
M_X = 3.0\times M_WB_1^{-\frac{1}{4}} = 2.6\times M_W\times 10^3\;\mbox{GeV} \approx 210\;\mbox{TeV}.
\end{equation}

As is seen from (\ref{math/22}), kaon LFV-decays may be sensitive to
very large energy scales, not accessible even at the future generation of 
supercolliders. However, models with the $s$-channel exchange of an $X^0$ particle are severely constrained, as they would result in a significant mass splitting between the $K^0_S$ and $K^0_L$ mesons due to the mixing process

\begin{equation}
\label{math/23}
\kXk,
\end{equation} 
shown in Fig. 4. The resulting $K^0_L - K^0_S$ mass difference 
is given by (see Refs. [25] and [28]):

\begin{equation}
\label{math/24}
\triangle m'_K \approx \frac{8}{3}m_Kf^2_K\left(\frac{h''}{M_X}\right)^2
 = \frac{8}{3}m_K(2\times 10^{-3}M_W)^2\left(\frac{h''}{M_X}\right)^2,
\end{equation}  
with $f_K \approx 160$ MeV $\approx 2 \times 10^{-3}M_W$.
Under the extreme assumption that process (\ref{math/23}) is responsible for
the entire kaon mass splitting $\triangle m_K$ (i.e. $\triangle m'_K =\triangle 
m_K = 3.49\times 10^{-12}$~MeV), we obtain from Eq. (\ref{math/24}):

\begin{equation}
\label{math/25}
\left(\frac{h''}{M_X}M_W\right)^2 = \frac{\triangle m_K}{m_K}
\frac{3}{8}\cdot \frac{10^6}{4} = 6.57\times 10^{-10}
\end{equation}
and therefore from Eqs. (\ref{math/20}) and (\ref{math/25}) we find

\begin{equation}
\label{math/26}
B_1 \leq 2.86\times 10^4\left(\frac{h'}{h''}\right)^2\left[\frac{h''}{M_X}M_W\right]^4 
 = 1.23\times 10^{-14}\left(\frac{h'}{h''}\right)^2,
\end{equation}
which is lowered even further, as one would expect $\triangle m'_K < \triangle m_K$, due to the fact that SM weak interactions alone can account for a significant fraction of $\triangle m_K$. The only possible way to make $B_1$ given by Eq. (22) comparable with the existing experimental limits (see Table 1) is to introduce large difference between the quark and lepton couplings of the $X^0$ boson, $(h'/h'') \gtrsim 10$. It should be noted that such a ratio of coupling constants might not be unnatural. For example, a similar ratio of the $W$-boson couplings to leptons and strange quarks is observed in the SM: $\frac{1}{\sin\vartheta_c} \sim 5$ (for heavier quarks this ratio is even larger). Nevertheless, bearing in mind that $\triangle m'_K$ is expected to 
be much smaller than $\triangle m_K$, it seems that the only plausible scenarios that would yield detectable values of $B_1$ are either if there is exists a symmetry that suppresses the $K^0\rightleftarrows X\rightleftarrows \bar{K}^0$ mixing, or if the decay (6) is due to  a different mechanism (e.g, an exchange of a leptoquark with cross-generational couplings, see Fig.~5; note that such a leptoquark exchange will also impact $\mu\to e$ conversion in $\mu^- + (A,Z)\to e^- + (A,Z)$ and other LFV-processes involving leptons and quarks).

\begin{figure}
\centering \includegraphics[height=6cm,width=9cm]{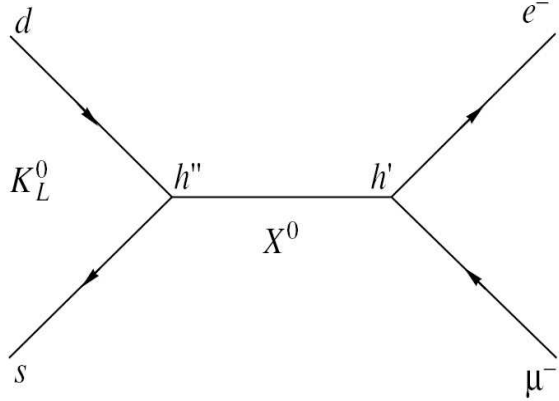}
\caption{Feynman diagram for the $K^0_L \to e^\mp \mu^\pm$ LFV-decay due
to an $s$-channel exchange of a heavy new boson $X^0$.}
\end{figure}
\begin{figure}
\centering \includegraphics[height=6cm,width=9cm]{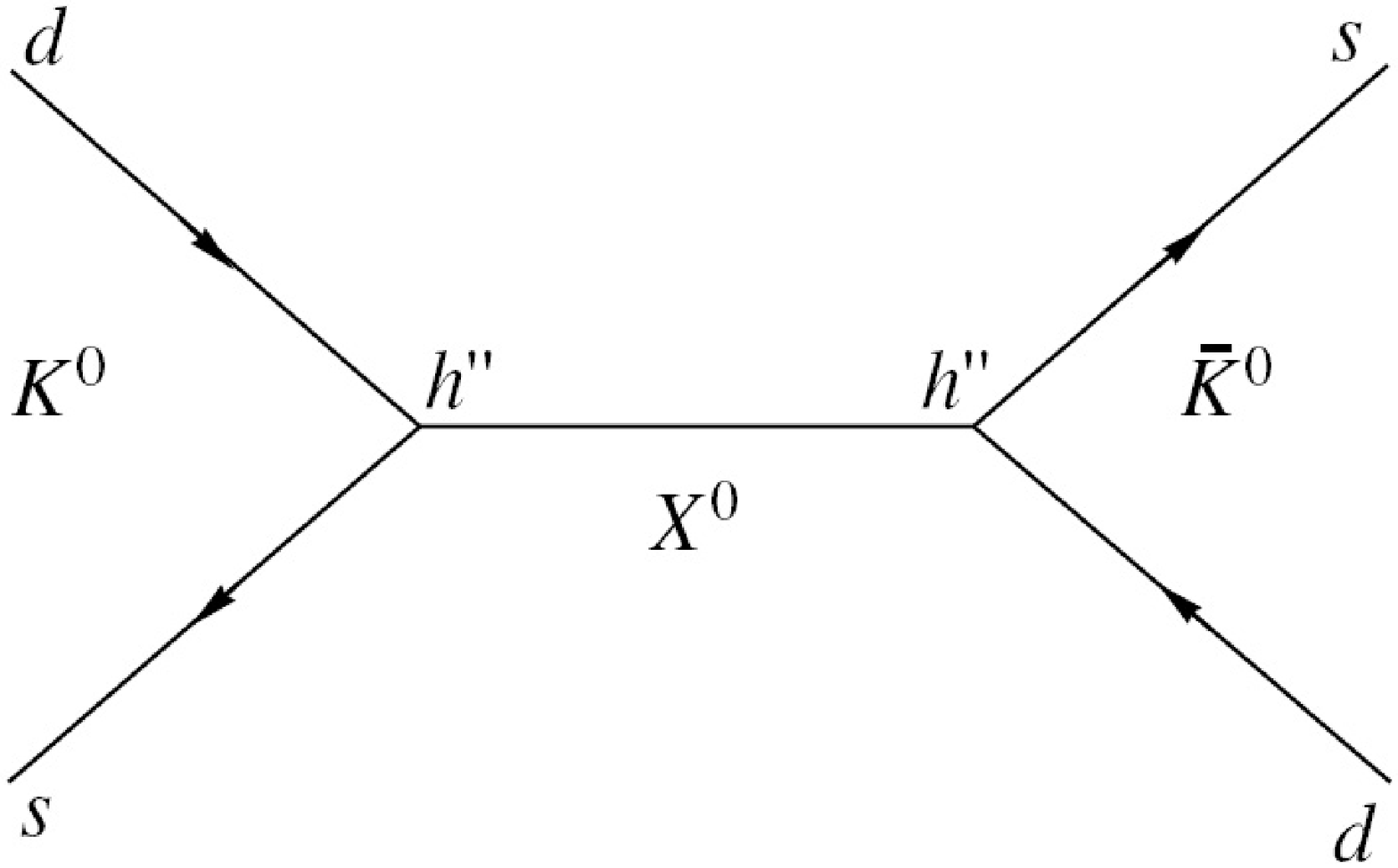}
\caption{The $\kXk$ mixing in models with the $X^0$-exchange.}
\end{figure}
\begin{figure}
\centering \includegraphics[height=6cm,width=9cm]{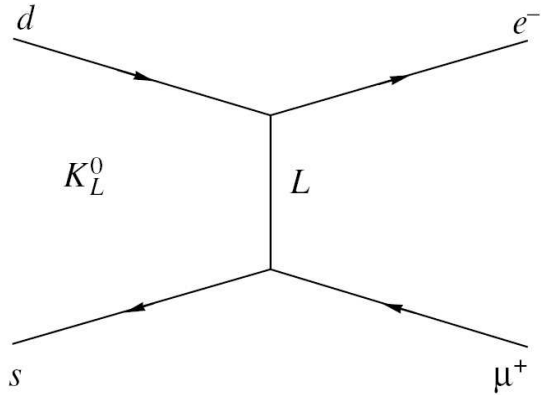}
\caption{Feynman diagram for LFV kaon decay via a $t$-channel 
exchange of a leptoquark with cross-generational couplings, L.}
\end{figure}

The pseudoscalar and axial-vector amplitudes for the $s\to d \bar{\mu} e^-$ process ($\klem$ decay), as well as the vector and scalar amplitudes for the $K\to\pi e\bar{\mu}$ decay with the $s$-channel $X^0$ boson exchange have been calculated in Ref. [28] assuming generic operators

\begin{equation}
\label{math/27}
\left.\begin{array}{cl}
Q_{V,A} = & \frac{g_X^2}{2M_X^2}\bar{d}\gamma_\alpha\left[C_{Lq}P_L + C_{Rq}P_R\right] s \bar{\mu}\gamma^\alpha
\left[C_{Ll}P_L + C_{Rl}P_R\right]e + {\rm h.c.}, \\
Q_{S,P} = & \frac{g_X^2}{2M_X^2}\bar{d}\left[C'_{Lq}P_L + C'_{Rq}P_R\right] s \bar{\mu}
\left[C'_{Ll}P_L + C'_{Rl}P_R\right]e + {\rm h.c.}
\end{array}\right.,
\end{equation}
where $P_L = (1-\gamma_5)/2$, $P_R = (1+\gamma_5)/2$, and $g^2_X$, $C$, $C'$ are the interaction constants.
The resulting branching fractions for the $K \to e \bar{\mu}$ and $K\to\pi e\bar{\mu}$ decays are shown 
in Fig.~6 as a function of the mass parameter $M_X$ for the case of the SM-like $X$-coupling ($g^2_X = g^2$)
and for $C = C' = 1$ (see Ref. [28]).

\section{Lepton-flavor violating kaon decays and fundamental fermion generations with approximately-conserved generation number}

As was mentioned above, searches for LFV in kaon decays are complementary to the rare muon-decay experiments looking for $\mu\to e\gamma$, $\mu\to 3e$, or $\mu^- + (A,Z)\to e^- + (A,Z)$. Kaon decays are the only processes 
sensitive to the neutral-current transition $s \to d \bar{\mu} e$, which may have special properties. These processes offer a unique possibility to study decays that violate fundamental-generation quantum number and to derive the selection rule for transitions involving this quantum number.

Indeed, as both quarks and leptons of different generations play role in these
LFV-decays, it may be possible to compensate a generation change in the quark 
sector by a corresponding change in the lepton sector. This possibility has
been discussed in the past and is best summarized in Ref.~\cite{30}, where a
a new quantum number $G$ was introduced to characterize fundamental fermion 
generations (1) of the SM. All flavor-changing neutral current (FCNC) 
LFV-processes are classified in the corresponding change in this quantum number 
between the initial and final states: $\triangle G = G_{\mbox{fs}} - G_{\mbox{is}}$.

If only the transitions between the fermions of the first two generations are 
considered, it is possible to assign arbitrary values of $G$ to these generations.
For example, one can assign $G_1 = 2$ for the fermions of the first generation and 
$G_2=1$ for those of the second generation (corresponding antifermions will have 
$G_1=-2$ and $G_2=-1$, respectively). Then, as is seen from the diagrams with the $X^0$-boson exchange shown in Fig.~7, various FCNC processes can be classified according to the change in $G$, $\triangle G$, as shown in Table~2.

\small
\begin{table}[t]
\caption{Classification of LFV-processes via the change in the generation quantum 
number $\triangle G$ in the model of Ref.~[30].}
\begin{tabular}[t]{c|c|l}\hline
Order in $\Delta G$ & $\triangle G$ &  Processes  \\ \hline 
First       & $\triangle G =0$  & $\kpm$; $\klem$; $K^0_L\to\pi^0e^\mp\mu^\pm$  \\ \hline 
Second      & $|\triangle G|=1$  &  $\mu\to 3e$; $\mu\to e\gamma$; $\mu^-+N\to e^- +N$ \\  \hline
Third       & $|\triangle G|=2$  & $\kk$ $(\triangle m(K^0_L-K^0_S))$; $\mu^-e^+\to\mu^+e^-$; $\kpme$ \\ \hline
\end{tabular}
Some examples to illustrate the $\Delta G$ selection rule:
\begin{itemize}
\item[a)] $\begin{array}[t]{cl}
\kpm \,\} & (\bar{s}u)\to (u\bar{d})\mu^+ e^- \\
      & G_{in} = -1+2=+1; G_{fin} = 0+(-1) + 2 = +1; \triangle G = G_{fin} - G_{in} =0 
\end{array}$
\item[b)] $\begin{array}[t]{l}
\begin{array}[t]{cl}
K^0\to e^-\mu^+ \, \} & (\bar{s}d)\to e^-\mu^+ \\
                    & G_{in} =+1; G_{fin} = +1; \triangle G = 0
\end{array} \\
K^0\to e^+\mu^- \, \}\, G_{in} = +1; G_{fin} = -1; \triangle G = -2 \\
\bar{K}^0\to e^-\mu^+\,\}\,\triangle G = +2 \\
\bar{K}^0\to e^+\mu^-\,\}\,\triangle G = 0 
\end{array}
\begin{array}[t]\}{l}.
\mbox{If the dominant decay has~} \triangle G = 0, \\[6pt]
K^0_L\simeq K^0_2 = \frac{1}{\sqrt{2}}|K^0 \,\,\,-\,\,\, \bar{K}^0\rangle\to e^\mp\mu^\pm \\
\qquad\hspace*{19mm}\tov e^-\mu^+\tov e^+\mu^-, \\
\mbox{i.e. there is an additional factor} \\
\mbox{of~}\frac{1}{2} \mbox{~in the matrix elements for} \\
\mbox{these decays}.
\end{array}$
\item[c)] $\begin{array}[t]{l}
\mu^-   + N\to e^- + N \\
                       \\                        
\end{array}
\begin{array}[t]\}{c}\}
G_{in} = +1 + G(N) \\
G_{fin} = +2+G(N) 
\end{array}
 \triangle G = +1$
\item[d)]  $\begin{array}[t].{r}\}
\kk \,\}\, G_{in} = +1 \\
 G_{fin} = -1 
\end{array}
 \triangle G = -2$
\end{itemize}
\end{table}
\normalsize

Physics behind the quantum number $G$ and the very existence of the $\triangle G = 0$ selection 
rule have not been established. Searching for LFV in kaon decays would only be interesting if this
selection rule holds (at least approximately) and thus suppresses the $|\Delta G| > 0$ processes 
significantly. If this is the case, the constraints on the $t$-channel exchange of the $X^0$ boson 
from the $K^0\rightleftarrows X^0\rightleftarrows \bar{K}^0$ mixing would be relieved, as the mixing 
occurs only at $|\Delta G| = 2$ (i.e., at the third order, see Table~2), which is suppressed 
significantly by the $\Delta G = 0$ selection rule. It should be also noted that models with the
approximately conserved generation number would result in significant difference between the $\kpm$ 
($\Delta G = 0$) and $\kpme$ ($|\Delta G| = 2$) decay rates (see Table~2).

\begin{figure}
\centering \includegraphics[height=4.5in,width=15cm]{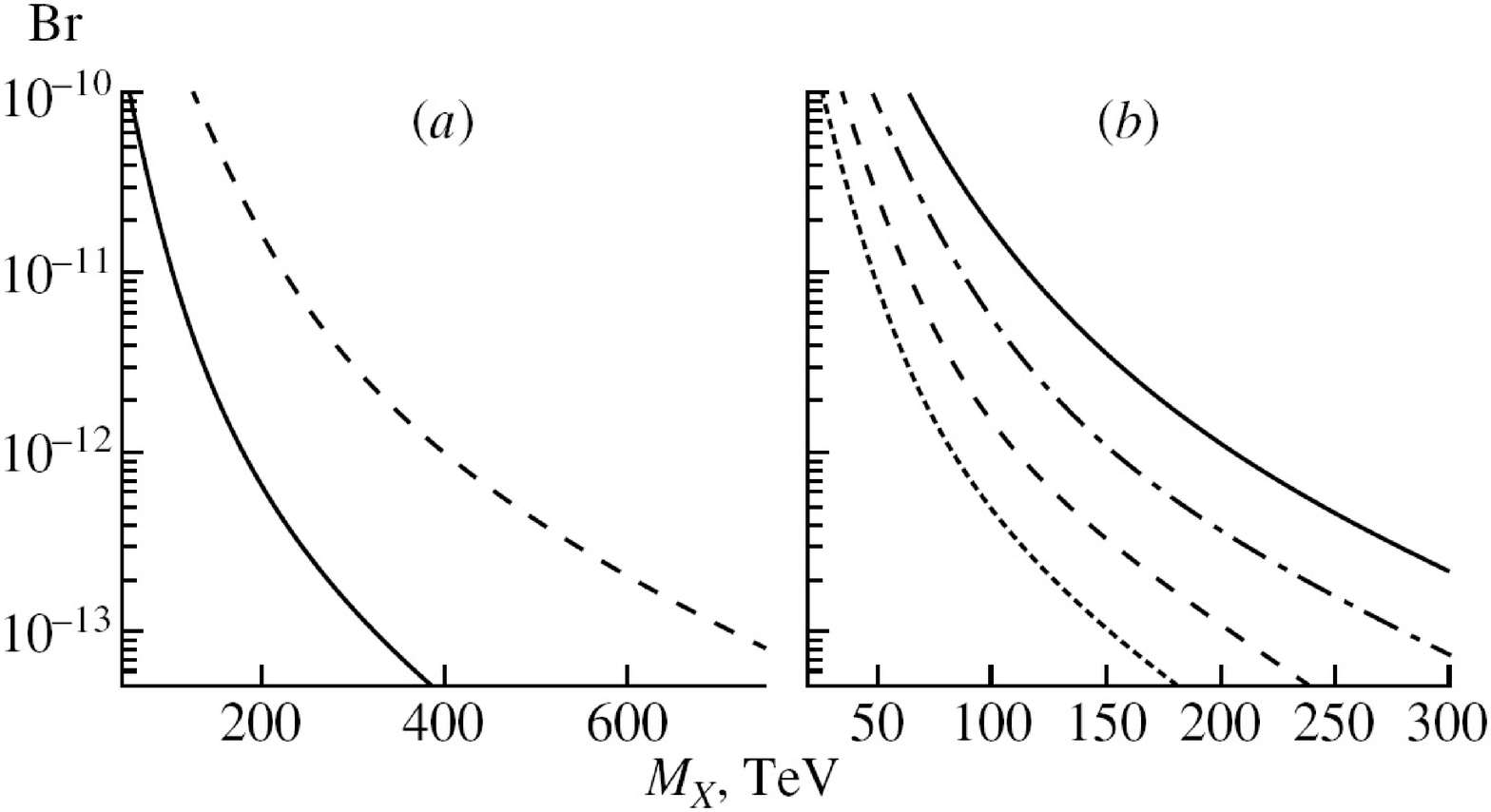}
\caption{Branching fractions for the LFV decays: (a) $K^0_L \to e \bar{\mu}$ and
(b) $K \to \pi e \bar{\mu}$, as a function of the mass parameter
$M_X$ in Eq. (27) (assuming SM-like $g_X = g$ and $C = C' = 1$).
In (a), the solid (dotted) curve corresponds to the $A$ ($P$) exchange.
In (b), the $K^0_L$ decay mode is represented by the solid ($S$) and 
dashed ($V$) curves, while the $K^+$ mode is represented by the 
dash-dotted ($S$) and dotted ($V$) curves (see Ref. [28] for detail). 
Here $V$, $A$, $S$, and $P$ correspond to vector, axial-vector, scalar, and 
pseudoscalar interactions.}
\end{figure}
\begin{figure}
\centering \includegraphics[height=15cm,width=16cm]{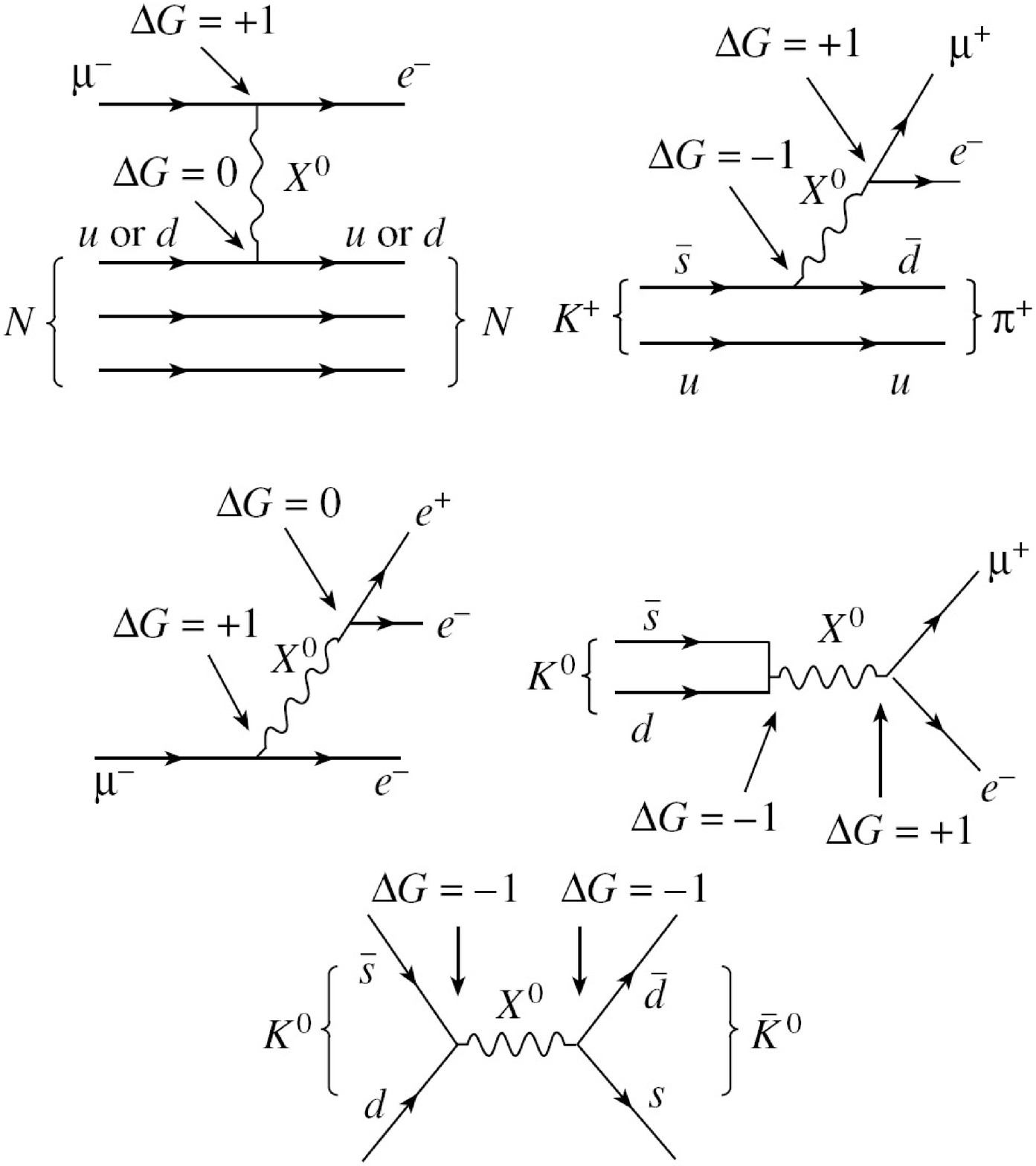}
\caption{Feynman diagrams for lepton-flavor violating processes in models
with the generation-number selection rule (see text and Table 2 for detail).}
\end{figure}

All in all, if the fermion generation quantum number $G$ is approximately 
conserved and the selection rule $\triangle G = 0$ holds, further searches for rare 
kaon LFV-decays may be of great importance.

Let us illustrate this statement by considering a dynamical model with extra spatial 
dimensions discussed in Refs. \cite{31} and especially \cite{32}. Listed below,
is a brief summary of the main aspects of this model:

\begin{itemize}
\item[a)] The model \cite{31,32} is based on a $M^4 \times S^2$ space-time with usual four-dimensional Minkowski 
space-time and two extra spatial dimensions compactified on a two-dimensional sphere with the radius $R$. The 
fundamental SM fermions form a single generation in the six-dimensional space-time and this single generation 
is then reduced to the three fundamental generations of the SM (see (1)) localized in different regions of the multidimensional space and characterized by the generation quantum numbers $G_i$. These quantum numbers correspond 
to quantized angular momentum in the compactified $S^2$ space and are fixed to $G_1=2$, $G_2=1$, and $G_3=0$ for the three SM fermion generations (and $G_1=-2$, $G_2=-1$, $G_3=0$ for the corresponding antifermions).

\item[b)] A distinct feature of the model is the existence of LFV-processes with charged leptons in the final 
state. The probabilities of these LFV-transitions are determined by the structure of extra space, 
i.e., by the inverse of the compactification radius $\frac{1}{R}$, corresponding to the energy spacing between 
the Kaluza-Klein modes of gauge bosons  propagating in extra dimensions.

\item[c)] As the mixing between different fermion generations is small, the generation quantum number 
is approximately conserved and the selection rule $\Delta G = 0$ applies. Thus, branching fractions 
for the $|\triangle G| > 0$ decays are suppressed:
\begin{equation}
\label{math/28}
\BR\sim  |\varepsilon^{|\triangle G|}|^2,
\end{equation}
where $\varepsilon$ is the generation mixing parameter.

\item[d)] The strength of fermion interactions with the SM gauge bosons and Higgs bosons 
is tuned (see~\cite{31,32}) to reproduce the observed fermion masses and CKM-matrix elements. 
From these data the mixing parameter for the quark sector $\varepsilon =\varepsilon_q$ is estimated 
to be $\varepsilon_q\sim 10^{-2}$. For the lepton sector the mixing parameter $\varepsilon=\varepsilon_L$ 
can not be extracted in a model-independent way. Therefore we will use two values of this parameter, 
which do not contradict the existing experimental data: a) $\varepsilon_L\simeq\varepsilon_q\simeq
 10^{-2}$ and b) $\varepsilon_L\simeq 10^{-3}$.
\end{itemize}

Following Ref. [32], we consider three types of FCNC LFV-processes with different values of 
$\triangle G$ and perform simple evaluation of their probabilities by comparing lepton-flavor 
violating and dominant, lepton-flavor conserving processes with similar kinematics.

\begin{itemize}
\item[1.] LFV decays, allowed by the selection rule $\triangle G=0$
 ($\klem$; $\kpm$).

The branching fraction of the decay $\klem$ can be compared with that for the SM 
decay $\kmn$, following procedure of Section~2 and taking into account
the factor of $\frac{1}{2}$ in the amplitude for the $\triangle G = 0$ decays
(see note in Table 2). For this decay in model [32]:

\begin{equation}
\label{math/29}
\Gamma(K_L^0 \to e^{\mp} \mu^{\pm}) = \frac{1}{16\pi}
\left(\frac{g^2}{16\cos^2\theta_W} \zeta R^2\right)^2 m_K m^2_{\mu}f^2_K
\left(1-\frac{m^2_{\mu}}{m^2_K}\right)^2,
\end{equation}
where $\zeta = 0.4$ is a parameter of the model. Therefore,
\begin{eqnarray}
\label{math/30}
B_1 & = & \frac{\Gamma(\klem)}{\Gamma(\kmn)} = 
\left[ \frac{\BR(\klem)}{\BR(\kmn)}\frac{\tau(K^+)}{\tau(K_L^0)} \right]
= \frac{\left(\frac{g^2}{16\cos^2\theta_W} \zeta R^2\right)^2}
{2 \times \left(\frac{g^2}{8M^2_W}\right)^2 \sin^2\theta_c} = 
\nonumber \\
& & \left(\frac{\zeta}{\sin^2\theta_c}\right)^2 R^4 
\left(\frac{M_W}{\cos\theta_W}\right)^4 \cdot \frac{1}{8} < 1.75 \times 10^{-12}
\end{eqnarray}
(see Eq. (18)). This can be translated in the following limit on the inverse radius of compactification in this model:
\begin{eqnarray}
\label{math/31}
\frac{1}{R} & > & \frac{M_W}{\cos\vartheta_W}
\!\left[\frac{\BR(\kmn)}{\BR(\klem)}\!\frac{\tau(K^0_L)}{\tau(K^+)}
\left(2\sin^2\vartheta_W - \sin^2\vartheta_W +\frac{1}{4}\right)\right]^{1/4}\!\!
\left(\frac{\zeta}{\sin\vartheta_c}\right)^{1/2}\!\!\!\!
\times 0.595 =
 \nonumber \\ & & 
\frac{M_W}{\cos\vartheta_W}\left(\frac{\zeta}{\sin\vartheta_c}\right)[\BR(\klem)^{-1}\times
 0.64\times 4.19]^{1/4} \times 0.595 =
\nonumber \\ & &
101\sqrt{\zeta}\; \mbox{TeV}\simeq 64\; \mbox{TeV}.
\end{eqnarray}

For another LFV-decay with $\triangle G = 0$, $K^+\to\pi^+\mu^+e^-$, the following 
limit has been derived in Ref. \cite{32}:
\begin{equation}
\label{math/32}
\frac{1}{R}  >  \frac{M_W}{\cos\vartheta_W}\left(\frac{\zeta}{2\sin\vartheta_c}\right)^{1/2}
 \left[\frac{\xi\BR(K^+\to\pi^0\mu^+\nu)}{\BR(K^+\to\pi^+\mu^+ e^-)}\right]^{1/4}.
\end{equation}
Here  $\xi = \left(4\sin^2\vartheta_W/3-1)^2(1+(4\sin^2\vartheta_W-1)^2\right) + 
(16\sin^2\vartheta_W\cos^2\vartheta_W/3)^2 \approx 1.38$. 
From the current limit $\BR(K^+\to\pi^+\mu^+ e^-) < 1.2 \times 10^{-11}$,
the following limit on the compactification scale has been obtained:
\begin{equation}
\frac{1}{R}>22\; \mbox{TeV}.
\end{equation}

\item[2.] LFV-suppressed decays with $|\triangle G|=1$.

First, we consider the decay $\mu^+\to e^+e^+e^-$, which proceed with
$|\triangle G|=1$ (see Table~2). It can be compared with the SM decay
$\mu^+\to e^+\nu_e\bar{\nu}_\mu$:
\begin{eqnarray}
\label{math/34}
\frac{\BR(\mu^+ \to 3e)}{\BR(\mu^+ \to e^+ \nu_e \bar{\nu_{\mu}})}
 & = & (M_WR^4)(\varepsilon_L)^2\zeta^2\left[\frac{1+20\sin^4\vartheta_W}
{2\cos^4\vartheta_W}\right] = \nonumber \\
& &  (M_WR^4)(\varepsilon_L)^2\zeta^2\times 1.90 < 1.0 \times 10^{-12},
\end{eqnarray}
which corresponds to 
\begin{equation}
\frac{1}{R}>60\; \mbox{TeV}\sqrt{\varepsilon_L} = \left\{\begin{array}{c}
6.0\;\mbox{TeV}\;(\varepsilon_L = 10^{-2}) \\
1.9\;\mbox{TeV}\;(\varepsilon_L = 10^{-3})\end{array}\right..
\end{equation}

The $\mu\to e\gamma$ decay in this model is suppressed additionally by a
loop factor and therefore has a small branching fraction compared to that in the $\mu \to 3 e$ decay.

Potentially the most interesting LFV-process with $|\triangle G|=1$ is 
neutrinoless $\mu\to e$ conversion in a field of a nucleus  $\mu^- + Z\to 
e^- + Z$. The dependence of the probability for this LFV-process on the 
compactification scale $R$ is given by (see Ref. [32]):

\begin{eqnarray}
\label{math/36}
F & = & \frac{\Gamma(\mu^- + Ti\to e^- +Ti)}{\Gamma(\mu^-+Ti\to \mbox{capture})}= \\
& & 2(\varepsilon_L)^2 \alpha^3_{QED} R^4
m_\mu^4\left[\frac{\zeta F(q^2)}{\pi}\right]^2Z_{\rm eff}^4\left[\frac{\kappa m_\mu}{Z
\Gamma(\mu^-\to \mbox{capture})}\right] M_W^4G_F^2, \nonumber 
\end{eqnarray}
where $\Gamma(\mu^-\to \mbox{capture}) = 
\Gamma(\mbox{capture}) = 2.6\times 10^6$ sec$^{-1} = 1.71 \times 10^{-15}$ MeV for the $Ti$ nucleus;
$Z=22$, $N=26$ are the numbers of protons and neutrons in the nucleus; the effective electric charge 
of the nucleus $Z_{\rm eff} = 17.6$; nuclear formfactor $|F(q^2)|\approx 0.54$; and $\kappa = 220$. 
Using Eq. (\ref{math/36}) and $F < 4.3 \times 10^{-12}$ leads to the following bound:
\begin{equation}
\label{math/37}
\frac{1}{R}  >  m_\mu Z_{\rm eff}\left[\frac{2\alpha^2_{QED}m_\mu |F(q^2)|^2\kappa}{\pi^2\Gamma_{capture}
Z F}\right]^{1/4} M_W G_F^{1/2}\zeta^{1/2}(\varepsilon_L)^{1/2},
\end{equation} 
or
\begin{equation}
\label{math/38}
\frac{1}{R}>78\; \mbox{TeV}(\varepsilon_L)^{1/2} = \left\{\begin{array}{c}
7.8\;\mbox{TeV}\;(\varepsilon_L = 0.01) \\
2.5\;\mbox{TeV}\;(\varepsilon_L = 0.001)\end{array}\right..
\end{equation}

\item[3.] $K^0\to X\to \bar{K}^0$ mixing with $|\triangle G|=2$.

As was shown in \cite{32}, the $K^0 \rightleftarrows  X \rightleftarrows \bar{K}^0$
transition is of the third order in $\triangle G$ and, consequently, is strongly 
suppressed by the selection rule (\ref{math/28}). Not surprisingly, the bounds on the 
compactification radius from this mixing process are very weak:
\begin{equation}
\label{math/39}
\mbox{a)} \qquad \frac{1}{R} < 1.5\;\mbox{TeV (from}\; \triangle m_K);\qquad\qquad\qquad\qquad
\qquad\qquad\qquad\qquad\qquad
\end{equation}
\begin{equation}
\label{math/40}
\mbox{b)} \qquad \frac{1}{R} < 2.6\,\mbox{TeV (from possible effect on the $CP$-violation parameter}\;\varepsilon_K).
\end{equation}
All in all, $K^0\rightleftarrows X\rightleftarrows \bar{K}^0$ mixing, being third order in $\triangle G$, imposes no restriction on the branching fraction of the LFV-decay $\klem$.
\end{itemize}

The ability to simultaneously accommodate several types of neutral-current LFV processes with different 
$|\triangle G|$ within the common framework of Ref. \cite{31,32} is particularly attractive. In this model, 
using Eqs. (\ref{math/30}) -- (\ref{math/38}), branching fractions of various LFV processes can be 
expressed as
\begin{equation}
\label{math/41}
\BR(LFV)_i = a_i/R^{-4}_{\mbox{eff}_i},
\end{equation}
with
\begin{equation}
\label{math/42}
R_{\mbox{eff}_i}^{-1} =[R^{-1}(\varepsilon^{\triangle G})^{1/2}]_i,
\end{equation}
as given by the selection rule (\ref{math/28}).

For processes allowed by the selection rule, i.e., the ones with $\triangle G=0$,  $R^{-1}_{\rm eff} = R^{-1}$.
For the $|\triangle G|=1$ LFV processes we consider two values of the lepton mixing parameter
$\varepsilon_L$: a) $\varepsilon_L\simeq\varepsilon_q\sim 10^{-2}$  and b) $\varepsilon_L \sim 10^{-3}$
(clearly, the $\triangle G$ selection rule is enforced more strongly for the latter case).

The branching fractions for LFV-decays can be expressed as
\begin{eqnarray}
\label{math/43}
\BR(LFV)_i = a_i/[R^{-4} (\varepsilon^{|\triangle G|})^{2}]_i,
\end{eqnarray}
which gives
\begin{eqnarray}
\label{math/44}
a_i = \BR(LFV)_i [R^{-4}(\varepsilon^{|\triangle G|})^2]_i.
\end{eqnarray}
Therefore, similar sensitivity to possible lepton flavor violation can be achieved in
different LFV-processes within the model~\cite{31,32}, as long as they have the similar 
values of the effective radius $R_{\mbox{eff}_i}^{-1} =[R^{-1}(\varepsilon^{|\triangle G|})^{1/2}]_i$.

From the bound on the compactification scale of $\frac{1}{R} = 64$~TeV, which comes from 
the most sensitive $\klem$ decays with the current bound of $\BR(\klem)\leq 4.7\times 10^{-12}$, it is possible 
to estimate the expected branching fractions for other LFV-processes with similar sensitivity 
to lepton flavor violation:
\begin{eqnarray}
\label{math/45}
\BR(LFV)_i|_{R^{-1}_{\rm eff}=64\;\rm TeV}  & = & \left\{\BR(LFV)_{\mbox{exp}_i} = \frac{a_i}{[R^{-4}
(\varepsilon^{\triangle G})^2]_i}\right\} \frac{[R^{-4}(\varepsilon^{\triangle G})^2]_i}
{(64\,\mbox{TeV})^4}  = \nonumber \\
& & a_i/(64\,\mbox{TeV})^4
\end{eqnarray}
Assuming that the sensitivity of $\BR(\klem) \sim 10^{-14}$ can be achieved in
future kaon experiments, which corresponds to $\frac{1}{R} = 298$~TeV, the following sensitivities
for other LFV-processes can be obtained:
\begin{equation}
\label{math/46}
\BR(LFV)_i|_{R^{-1}_{\rm eff}=298\;\rm TeV} = a_i/(298\,\mbox{TeV})^4.
\end{equation}
We summarize the results from Eqs. (45) and (46) in Table~3. As is seen from the Table,
in the model with two extra dimensions~\cite{31,32} the sensitivity of kaon decays with
$\triangle G=0$, especially $\klem$, to lepton flavor violation may be several orders of 
magnitude larger than that in the LFV muon decays with $|\triangle G|=1$, due to 
significant suppression of $|\Delta G| > 0$ transitions.
 
It should be noted that the discussed model~\cite{31,32} is just one of many possible models with lepton flavor 
violation, as other, more complicated LFV-mechanisms may exist. In some of these models,
purely leptonic processes ($\mu\to 3e$, $\mu\to e\gamma$) are more sensitive to LFV compared with 
the quark-lepton processes; in others the latter are more sensitive than the former. It is 
important to study and compare various types of quark-lepton processes, e.g. $s\to d\bar{\mu} e$ and 
$d\to d\bar{\mu}e$, as their sensitivities to LFV may differ within different 
theoretical frameworks and depend on the details of a particular model (e.g., the 
leptoquark charge, or a particular flavor of a technicolor model, etc).
Finally, processes with lepton number violation, such as neutrinoless
double $\beta$-decay, $Z \to (Z+2) + 2e^-$, or exotic decays, such as $K^+\to\pi^-\mu^+\mu^+$,
may play an important role in understanding lepton flavor violation.

To conclude, the problem of lepton flavor violation is of crucial interest due to its possible sensitivity 
to very large energy scales, well beyond the reach of the next generation of supercolliders (see 
\cite{01,11,25,26,27,28,30,31,32,33,34,35,36,37,38,39,40,41,42,43,44,45} and references therein). 
An ambitious program of new searches for lepton flavor violation in rare muon decays is 
currently being developed. It is planned to increase the sensitivity of these experiments by 3 to 5 orders 
of magnitude compared with that achieved to date (see Table~1). While this program represents
an important direction for future studies of LFV, new searches for LFV in kaon decays are also important.
They are complementary to the muon-decay experiments and therefore should be considered as an independent 
direction within the general program of understanding the mechanism of lepton flavor violation. We discuss
several possible future kaon experiments in the next section.

\section{Future prospective searches for LFV in kaon decays}

\begin{figure}
\centering \includegraphics[height=5in,width=10cm]{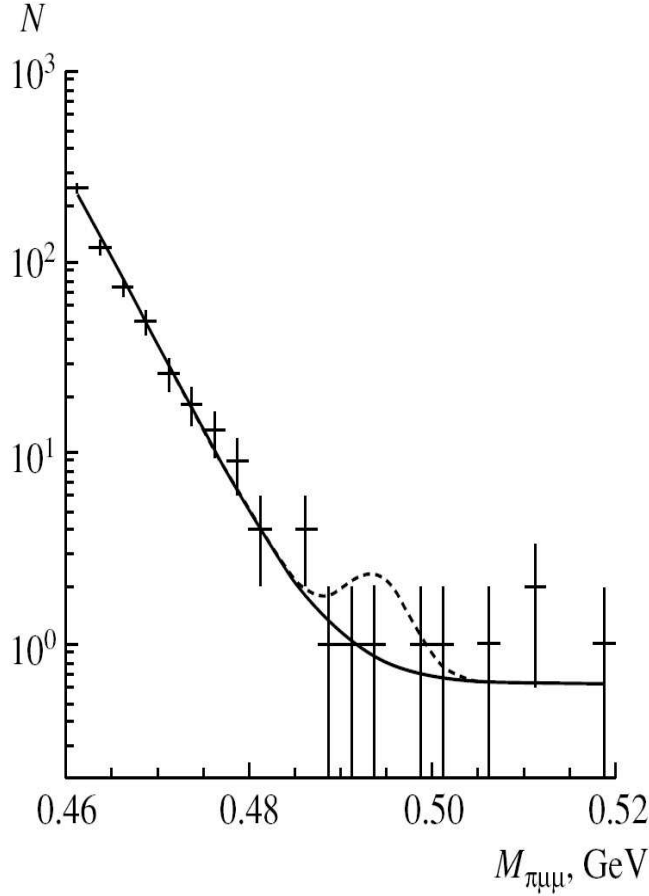}
\caption{E865 data for $K^+ \to \pi^- \mu^+ \mu^+$. The dashed curve
corresponds to $\BR(K^+ \to \pi^- \mu^+ \mu^+) \simeq 3 \times 10^{-9}$. 
From Ref. [6].}
\end{figure}

Future studies of rare kaon LFV-decays are possible if the following two main conditions are met:
\begin{itemize}
\item[a)] Building intensive kaon sources that would allow for significant increase in statistical sensitivity to rare decays. 
\item[b)] Further development of reliable modern methods of rare decay identification and suppression of background processes, which often limit the sensitivity of the existing experiments.
\end{itemize}

Both of these conditions can be met in a new generation of kaon experiments at the existing 
intermediate-energy (25--120~GeV) accelerators, as well as at the machines currently under construction.
The detectors for these new experiments must be based on modern technology to maximize their ability 
to operate at high rates, with the efficient identification of the decay products and very accurate momentum 
and time resolutions.

\begin{figure}[h]
\centering \includegraphics[height=12cm,width=12cm]{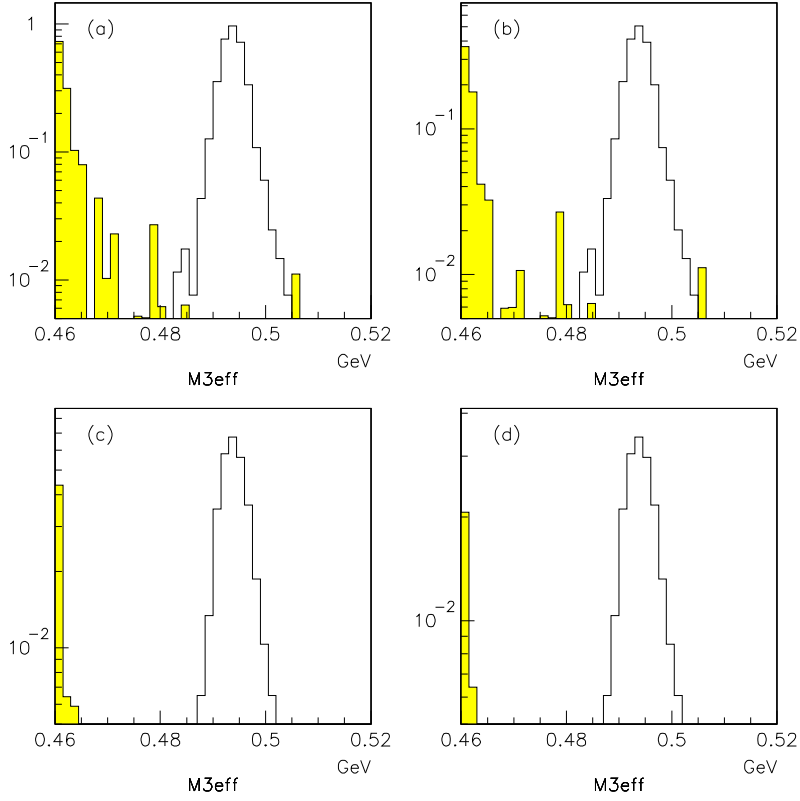}
\caption{Simulation of background to the $K^+ \to \pi^- \mu^+ \mu^+$
decay from the $K^+ \to \pi^- \pi^+ \pi^+$ process with two $\pi^{+} \to 
\mu^{+} \nu_\mu$ decays in flight in the CKM experiment [46,47] 
(shaded histograms). Open histograms that peak in each figure
correspond to the signal ($K^+ \to \pi^- \mu^+ \mu^+$), with different 
vertical scales. In a) ,b) background is suppressed by standard 
kinematic cuts similar to those used in the E865 experiment. In c) ,d)
background is further suppressed by special procedure based on redundant
muon momentum measurements in both the magnetic spectrometer and RICH-based
velocity spectrometer (see text and Refs. [46,47]). Comparison of the data
in this figure with those in Fig.~8 demonstrates that the double muon 
constraint in CKM can reduce background from the $\pi \to \mu \nu_\mu$ decays in
flight by over two orders of magnitude (to the level of $\BR(K^+ \to 
\pi^- \mu^+ \mu^+) \sim 10^{-11}$) compared to that in E865.}
\end{figure}

\small
\begin{landscape}
\begin{table}[c]
\caption{Sensitivities to various LFV-processes in the extra-dimensional model with approximately conserved fermion-generation quantum numbers \cite{32}.}
\begin{tabular}[t]{|p{0.17\hsize}|p{0.033\hsize}|p{0.12\hsize}|p{0.11\hsize}|p{0.09\hsize}
|p{0.15\hsize}|p{0.15\hsize}|p{0.10\hsize}|}\hline
 BR(LFV)~(90\% C.L.) (existing experimental data, see Table~1) &  
$|\triangle G|$ & $[R^{-1}(\varepsilon^{|\triangle G|})^{1/2}]_i$ (TeV) & 
$[R^{-4}(\varepsilon^{|\triangle G|})^2]_i$ (TeV)$^4$ & $a_i$(LFV) \ \ \ \ \ \ (TeV)$^4$ &
Expected branching fractions for the LFV-processes normalized to the existing 
 $\klem$ decay data and $R^{-1}=64$~TeV (see Eq. (54)) & Expected branching fractions for LFV-processes normalized to the
expected future sensitivity for the $\klem$ decay and $R^{-1} = 298$~TeV  (see Eq. (55)) & 
Expected sensitivities to the branching fractions of LFV-processes in future experiments\\ \hline
& & & & & & & \\
$\BR(\klem)<4.7\times 10^{-12}$ & 0 & 64 & $1.68\times 10^7$ & $7.90\times 10^{-5}$ & 
 $\BR(\klem)<$ \parbox{25mm}{$4.7\times 10^{-12}$} $R^{-1} = 64$ 
& $\BR(\klem) \sim 10^{-14}$ \ \ \ \ \ \ \ \ \ \ \ \ \ \  $R^{-1}=298$  & $10^{-13} - 10^{-14}$  \\ \hline
$\!\BR(K^+\!\to\!\pi^+\!\mu^+\!e^-)\!<1.2\times 10^{-11}$ & 0 & 22 & $2.45\times 10^5$ & $6.85\times 10^{-6}$& 
$4.08\times 10^{-13}$ & $0.87\times 10^{-15}$ & $10^{-12}$ \  \  (CKM)~[51] \\ \hline
$\BR(\mu^-\!\to\! e^+e^-e^-)<1.0\times 10^{-12}$ & 1 & 6.07$(\varepsilon_L \!=\! 10^{-2})$ 1.92$(\varepsilon_L\!=\! 
10^{-3})$ & 
\parbox{25mm}{$1.36\times 10^3$} $1.36\times 10$  & $1.36\times 10^{-9}$ $1.36\times 10^{-11}$ &
\parbox{25mm}{$0.81\times 10^{-16}$}  \ \ \ \ \ \ 
$0.81\times 10^{-18}$ & \parbox{25mm}{$0.17\times 10^{-18}$} \  \ \ \ \ \ $0.17\times 10^{-20}$ &
$10^{-14} - 10^{-15}$ [29] \\ \hline
$F = \frac{\Gamma(\mu^-\to e^-)}{\Gamma(\mu\,\mbox{capture})}<4.3\times 10^{-12}$ & 1 & 
7.80$(\varepsilon_L \!=\! 10^{-2})$ 2.47$(\varepsilon_L \!=\! 10^{-3})$ &
\parbox{25mm}{$3.70\times 10^3$} $3.70\times 10$ & $1.59\times 10^{-8}$ $1.59\times 10^{-10}$ &
\parbox{25mm}{$0.94\times 10^{-15}$}  \ \ \ \ \ \ \  
$0.94\times 10^{-17}$ & \parbox{25mm}{$0.20\times 10^{-17}$} \ \ \ \ \ \ \ $0.20\times 10^{-19}$ &
\parbox{20mm}{$\sim~10^{-17}$} (MECO)~[29] \\ \hline
\end{tabular}
Note that columns 6--8 list expected branching fractions for which the sensitivity to LFV processes will
be comparable to that in the $\klem$ (columns 6, 7) data or in proposed future $\klem$ experiments (column 8).
\end{table}
\end{landscape}
\normalsize

Although no specific proposals for future LFV kaon-decay experiments exist to date, certain 
studies in this area are underway \cite{37,46,47}. In the CKM experiment \cite{46}, 
along with the main goal to perform precision measurement of the rare decay 
$K^+\to\pi^+\nu\bar{\nu}$, it is planned also to do parallel studies of several kaon LFV-decays:
$K^+\to\pi^+\mu^+e^-$, $K^+\to\pi^+\mu^-e^+$, and $K^+\to\pi^-l^+l^+$ (with the emphasis on the unique process  
$K^+\to\pi^-\mu^+\mu^+$). It was shown that the superior design of 
the CKM detector would allow searches for LFV-decays to reach sensitivity at the level of $\BR\sim 10^{-12}$, i.e.,
an order of magnitude better than that achieved in the E865 experiment \cite{05,06}, see Table~1. (Here and in the rest of this section, we quote sensitivities based on statistical-only uncertainties; the sensitivities will be revised downward, once a full-blown analysis of systematic uncertainties is completed.)
Despite the lack of a detailed background analysis to the LFV searches performed in CKM thus far, 
some studies of the $K^+\to\pi^-\mu^+\mu^+$ decay have been carried out, with the results expected to be quite promising.

Note that the limit obtained by E865 [6] in this LFV-decay channel ($\BR <3 \times 10^{-9}$) was among the least restrictive limits set by this experiment, primarily due to copious background from the $K^+\to\pi^-\pi^+\pi^+$ 
decay with two $\pi^+$ mesons decaying in flight ($\pi^+\to\mu^+\nu_\mu$) and thus ``faking'' muons in the E865 detector. The results of this measurement are presented in Fig.~8. As was shown via full GEANT Monte Carlo simulation of the CKM detector, a standard kinematic analysis (similar to that done in E865\cite{13});) should yield a similar limit 
($\BR < 2.5\times 10^{-9}$, at the 90\% C.L.). However, the superior design of the CKM detector would
allow to perform an optimized analysis based on the redundant momentum measurements for one of the muons both 
in the magnetic spectrometer and RICH-based velocity spectrometer. Monte Carlo simulations that incorporate this additional measurement demonstrate suppression of the $\pi^+\to\mu^+\nu_\mu$ background by two orders of magnitude 
compared to the conventional technique, which would allow to reach sensitivity of $\BR(\kpmm) < 10^{-11}$ 
(see Fig.~9 and Ref. \cite{47} for detail).

While the CKM experiment at Fermilab has been initially approved, its current fate is unclear, due to the lack
of available funds. The CKM group is working now on descoping the apparatus in order to reduce its price 
and get the appropriate funding \cite{48}. We hope that it would still be possible to study LFV-decays, 
particularly the $K^+\to\pi^-\mu^+\mu^+$ decay, with the descoped detector.

While it's important to pursue further studies of a variety of kaon LFV-decays, the main priority 
should be given to a new $\klem$ measurement with maximum possible sensitivity. Below, we estimate
the sensitivity, which can be achieved in future searches, and discuss various ways of handling backgrounds
(see also \cite{37,49,50}). Table~4 summarizes sensitivity that can be achieved using the $K^0_L$-beam 
at the Fermilab Main Injector with one of the versions of the proposed KAMI experiment \cite{49,50} 
(KAMI NEAR \cite{49}). As is seen from the table, the sensitivity of this measurement is around
$\BR(\klem)\sim 10^{-14}$.

\small
\begin{table}[t]
\caption{The KAMI-NEAR project [49] with the $K^0_L$-beam at the Fermilab Main Injector.}
\begin{tabular}[t]{|p{0.31\hsize}|p{0.31\hsize}|p{0.31\hsize}|}

\hline Proton and secondary neutral beams & Decay volume for the $K^0_L$ decays
& Sensitivity to the $\klem$ decay \\
\hline
\parbox[t]{50mm}{Main Injector proton beam:}
\parbox[t]{50mm}{$E_p = 120$~GeV.}
\parbox[t]{50mm}{$I_p \simeq 3\times 10^{13}$p/cycle $\to\quad 3.6\times 10^{16}$p/hour.}
\parbox[t]{50mm}{ $\vartheta_{K^0_L}\simeq 8$mrad.}
\parbox[t]{50mm}{Beam acceptance} 
\parbox[t]{52mm}{$d\Omega$$=(2.5$mrad.)$\times(2.5$mrad).}
\parbox[t]{50mm}{$= 6.3\mu$ster.}
\parbox[t]{50mm}{$K^0_L$ flux on the target $1.5\times 10^9\,K^0_L/$cycle.}
\parbox[t]{50mm}{The ratio of intensities $n/K^0_L\simeq 20$.}  &
\parbox[t]{50mm}{The distance from the target to the decay volume: $L=40$m.}
\parbox[t]{50mm}{The decay length in the decay volume $Z=23$m.}
\parbox[t]{50mm}{The $K^0_L$ decay probability in this volume $\simeq 10\%$.}
\parbox[t]{50mm}{The number of $K^0_L$ decays  $1.2\times 10^8K^0_L$/cycle $\to 1.4\times 10^{11}K^0_L$/hour.}
\parbox[t]{50mm}{$\langle P(K^0_L)\rangle_{decay}\simeq 15$ GeV/c.}  &
\parbox[t]{50mm}{The number of the $K^0_L$ decays in 2 years of running (assuming 
 50\% of dead time):}
\parbox[t]{50mm}{$N=1.7\times 10^{15}\,K^0_L$ decays.}
\parbox[t]{50mm}{The $\klem\quad\varepsilon\sim 10\%$} 
\parbox[t]{50mm}{detection efficiency.}
\parbox[t]{50mm}{Sensitivity:}
\parbox[t]{50mm}{$\BR(\klem)\lesssim 10^{-14}$.} \\ 
& & In the KAMI experiment [50] with the reduced intensity the sensitivity is $\BR(\klem)\lesssim 10^{-13}$. \\ \hline
\end{tabular}
\end{table}
\normalsize

At the 70~GeV IHEP accelerator in Protvino the sensitivity of $\BR(\klem)\lesssim 10^{-13}$ can be achieved,
primarily due to the lower intensity of the proton beam and shorter run duration. Nevertheless, 
it is important to explore the possibility to increase the detector acceptance and the intensity of the 
neutral kaon beam by using asymmetric and large beam spot. For example, the use of such an asymmetric beam
in the KOPIO experiment at BNL \cite{51}, results in the acceptance increase from $\sim 6\times\mu$ster 
(see Table 4) to $\sim 500\times\mu$ster. The dimensions of the beam spot are $\sim 120\times 10$ cm$^2$ 
(see Fig.~10). Such a geometry would reduce efficiency losses due to the presence of a beam-hole in the 
detector and lower background from random coincidences.

Design characteristics of the proposed next generation of intermediate-energy accelerators (Fermilab's Proton Driver to increase the intensity of the Main Injector, J-PARC project at KEK, accelerators for neutrino factories, see Refs. [37,52]) would allow to achieve sensitivity to the rare $\klem$ decay better than $\BR\sim 10^{-15}$, limited mainly by background.

Clearly, a full-fledged proposal for a new $\klem$ measurement with the sensitivity at the level of 
$\BR\sim 10^{-13} - 10^{-14}$ would require careful studies of measurement technique and the detector 
design, detailed GEANT-based simulation of the detector response to both signal and background, 
and calibration studies. Here, we would like to point out just a few main design requirements for such 
a future experiment.

\begin{itemize}
\item[1.] Neutral kaon beam must be of an exceptional quality with minimal halo (at the level $<10^{-4}$ of the beam intensity). To achieve this, beam formation and the collimation system must be designed very carefully.
To illustrate this we present in Fig.~10 the results of a simulation of neutral beams for the KAMI
experiment at Fermilab \cite{50}, based on data for the $K^0_L$-beam for the KTeV experiment
and the beam proposed for use in KOPIO~\cite{51}. The neutral beam must be kept in high vacuum 
and should not interact with the detector material.  

\item[2.] Forward part of the detector must contain two magnetic spectrometers following the decay volume, with exceptional spatial and momentum resolution to achieve a redundant double momentum measurement necessary to suppress the background from $\pi \to \mu \nu$ decays. The vacuum chamber must be located inside the yokes of the magnets. Low-mass drift straw tubes (which can work in vacuum) should be used as tracking detectors. Such a system has been developed and tested for the CKM experiment and demonstrated to work in high vacuum \cite{53}. The primary beam must be channeled through a hole in the spectrometers; thus the counting rates in the detector will be mainly due to kaon decay products.

\item[3.] Muon and electron identification system must be located behind the spectrometers. The beam is channeled through this system via a thin evacuated beam-pipe. A transition radiation detector followed by an EM calorimeter (as in KTeV experiment) or a low-pressure Cherenkov detector can be used for electron identification. A conventional spectrometer can be used for muon identification (as in E871~\cite{02}). However, it might be advantageous to perform simultaneous electron and muon identification with a high-precision RICH counter, as was proposed for CKM. It has been demonstrated that an additional measurement of muon velocity in such a detector will reduce the dominant $\pi \to \mu \nu$ background (see the discussion on the $K^+\to\pi^-\mu^+\mu^+$ decay and Fig.~9).
\end{itemize}

Main background processes to the $\klem$ decay have been considered in Refs. \cite{02,27}.
The three main sources of background are:
\begin{itemize}
\item[a)] $K^0_L\to\pi^\pm e^\mp \nu_e$ decay with a soft neutrino, followed by the pion decay
$\pi^\pm\to\mu^\pm\nu_e$. In this case, the maximum mass of the $e^\mp\mu^\pm$ system is shifted noticeably
relative to the $M(K^0_L)$: $M(e^\mp\mu^\pm)_{max} = M(K^0_L) - 8.43$~MeV \cite{27}. Nevertheless, at the desired level of precision, even small non-Gaussian tails in the momentum resolution could result in contamination of the signal region from this process.
\item[b)] Misidentification of secondary particles leading to a shift in the invariant mass 
distributions.
\item[c)] Random coincidences of the decay particles from two different $K^0_L$-mesons. The rate of this accidental background is increased with the intensity of the beam (the latter is essential for achieving ultimate sensitivity).
\end{itemize}

For a proposal for a future $\klem$ experiment the suppression of all three background components 
must be studied very carefully with full GEANT-based simulation of the detector, perhaps
taking into account detector calibration. Particular attention must be devoted to non-Gaussian 
tails in the reconstructed invariant mass in the final states due to the $\pi\to\mu\nu_\mu$ decays 
or beam interactions in the detector. To reduce these effects it is important to minimize the 
amount of material in the detector and to develop highly-efficient tracking code using modern techniques
(see, e.g., \cite{54}).

To reduce the accidental background it is important to optimize the profile of the neutral beam (e.g., by using a distributed beam with an asymmetric profile, as in KOPIO \cite{51}, see Fig.~10). The readout system must be based on modern fast electronics. It may be possible to use novel high-rate tracking detectors based on Micromega TPC 
technology~\cite{53} as a particle-identification component of the detector, similar to the KABES detectors in NA48 at CERN. It might be worth considering a more complicated detection scheme for a two-body decay (such as $\klem$) experiment with a focusing superconductive solenoid with the magnetic field directed along the neutral kaon beam. However, such a system would 
complicate the detector design quite significantly.

It must be bear in mind, that if a new high sensitivity program for the search for $\klem$ decay will be
achievd, than simultaneously a very important information on $K^0_L\to\mu^+\mu^-$ decay and
form factor $F(q^2_1; q^2_2)$ of  $K^0_L\to\gamma^*\gamma^*$ vertex from  $K^0_L\to\mu^+\mu^- e^+e^-$
decay would   be obtained.

\begin{figure}
\centering \includegraphics[height=16cm,width=12cm]{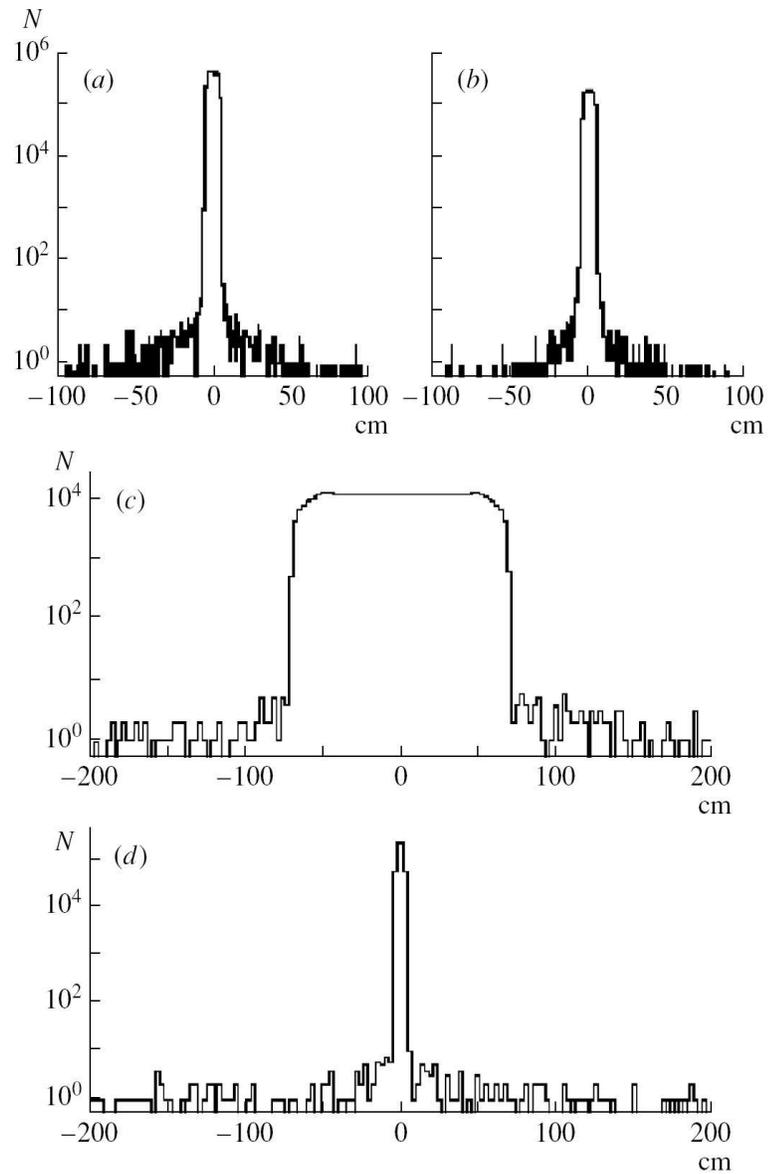}
\caption{Simulation of the profile for a low-halo neutral-kaon beam for KAMI [50] 
and KOPIO [51] experiments: a) $K^0_L$ beam in KAMI; b) Neutron beam in KAMI;
c) Horizontal profile of the neutral beam in KOPIO;
d) Vertical profile of the neutral beam in KOPIO.
Note that the neutral beam in KOPIO is significantly asymmetric in the vertical and horizontal directions (by $\sim 500 ~\mu ster$).}
\end{figure}

\section{Conclusions}

The unique opportunities in the LFV kaon decays $s\to de\bar{\mu}$ in models with approximate 
conservation of the
fundamental-generation quantum number ($\triangle G=0$) present a defensible physics case for a new generation of 
rare kaon-decay experiments at the existing and future intermediate-energy accelerators with high-intensity kaon 
beams, particularly for a high-precision measurement of the $\klem$ decay. The goal of these experiments is to 
increase the sensitivity to LFV kaon decays by 2--3 orders of magnitude compared to that achieved to date (see 
Table~1). These exciting, albeit difficult experiments are complementary to the new generation of rare LFV muon-decay experiments and therefore should be considered as an important part of general program of searches for lepton flavor violation in the processes with charged leptons.

\section{Acknowledgments}

I am grateful to M.V.~Libanov, E.Yu.~Nugaev, and S.V.~Troitsky for a number of stimulating discussions 
on lepton flavor violation. I would also like to thank the attendees of the Chicago Flavor Seminar for 
helpful remarks, and A.V.~Artamonov for his help in preparation of this article.
My special thanks to L.B.~Okun and G.~Landsberg for their help with edition of English text.

\end{document}